\documentstyle[12pt]{article}

\def\thefootnote{\fnsymbol{footnote}}
\def\bea{\begin{eqnarray}}
\def\eea{\end{eqnarray}}
\def\beq{\begin{equation}}
\def\eeq{\end{equation}}

\def\tH{{\tilde H}}
\def\tF{{\tilde F}}
\def\notD{\not{\hspace{-.05in}D}}
\def\LLL{\sum_\Phi{\lambda_\Phi\Lambda_\Phi^2\over 32\pi^2}}
\def\lll{\sum_P{\lambda_P\over 32\pi^2}}

\def\bmu{\bar{\mu}}
\def\[{\left [}
\def\]{\right ]}
\def\({\left (}
\def\){\right )}

\def\pp{\partial}
\def\M{\bar{M}}

\def\z{\bar{z}}

\def\STr{{\rm STr}}
\def\Tr{{\rm Tr}}

\def\K{{\cal K}}

\def\f{\bar{f}}

\def\L{{\cal L}}
\def\D{{\cal D}}

\def\hD{\hat{D}}
\def\hG{\hat{G}}
\def\hV{\hat{V}}

\def\m{\bar{m}}

\def\cM{{\cal{M}}}

\def\A{\bar{A}}

\def\Z{{\bar{Z}}}

\def\bb{\bar{b}}

\def\bth{\bar{\theta}}
\def\bv{\bar{\varphi}}

\def\bF{\bar{F}}

\begin{document}

\begin{titlepage}
\begin{center}

\hfill LBL-36051 \\
\hfill UCB-PTH-94/23 \\
\hfill August 1994 \\
\hfill hep-th/9408149 \\[.3in]

{\large \bf PAULI-VILLARS REGULARIZATION OF SUPERGRAVITY COUPLED TO CHIRAL AND
YANG-MILLS MATTER}\footnote{This
work was supported in part by
the Director, Office of Energy Research, Office of High Energy and Nuclear
Physics, Division of High Energy Physics of the U.S. Department of Energy under
Contract DE-AC03-76SF00098 and in part by the National Science Foundation under
grant PHY--90--21139.} \\[.2in]

Mary K. Gaillard \\[.1in]

{\em Department of Physics and Theoretical Physics Group,
 Lawrence Berkeley Laboratory,
 University of California, Berkeley, California 94720}\\[.5in]

\end{center}

\begin{abstract}

It is shown that the one-loop quadratic divergences of standard supergravity
can be regulated by the introduction of heavy Pauli-Villars fields belonging to
chiral and abelian gauge multiplets.  The resulting one-loop correction can be
interpreted as a renormalization of the K\"ahler potential.  Regularization of
the dilaton couplings to the Yang-Mills sector requires special care, and may
shed some light on chiral/linear multiplet duality of the dilaton
supermultiplet.

\end{abstract}
\end{titlepage}

\newpage
\renewcommand{\thepage}{\arabic{page}}
\setcounter{page}{1}
\def\thefootnote{\arabic{footnote}}
\setcounter{footnote}{0}

In extracting the phenomenological implications of an underlying supergravity
theory, the quadratic divergences arising at one loop have often been
considered~\cite{quad}.  However, the coefficients of the
quadratically divergent terms are unreliable in the absence of a manifestly
supersymmetric regularization procedure~\cite{sigma},~\cite{casimir}.  The
purpose of this Letter is to describe such a procedure.

The one-loop effective action $S_1$ is obtained from the term quadratic in
quantum fields when the Lagrangian is expanded about an arbitrary background:
\bea \L_{quad}(\Phi,\Theta,c) &=& -{1\over 2}\Phi^TZ^\Phi\(\hD^2_\Phi +
H_\Phi\)
\Phi + {1\over 2}\bar{\Theta}Z^\Theta\(i\notD_\Theta - M_\Theta\)\Theta
\nonumber \\ & & + {1\over 2}\bar{c} Z^c\(\hD^2_c + H_c\)c + O(\psi), \eea
where the column vectors $\Phi,\Theta,c$ represent quantum bosons, fermions
and ghost fields, respectively, and $\psi$ represents background fermions that
we shall set to zero throughout this paper. The fermion sector $\Theta$
includes a C-odd Majorana auxiliary field
$\alpha$ that is introduced to implement the gravitino gauge fixing condition.
The full gauge fixing procedure used here is described in detail
in~\cite{us},~\cite{us2}. Then the one loop bosonic action is given by
\bea S_1 &=& {i\over 2}\Tr\ln\(\hD_\Phi^2 + H_\Phi\)
-{i\over 2}\Tr\ln\(-i\notD_\Theta + M_\Theta\)
+ {i\over2}\STr\ln\(\hD_c^2 + H_c\)
\nonumber \\ &=& {i\over 2}\STr\ln\(\hD^2 + H\) + T_-, \eea
where $T_-$ is the helicity-odd fermion contribution which contains no
quadratic divergences, and the helicity-even contribution is determined by
\beq \hD^2_\Theta + H_\Theta \equiv
\(-i\notD_\Theta + M_\Theta\)\(i\notD_\Theta + M_\Theta\).\eeq
The field-dependent matrices $H(\phi)$ and $\hD_\mu(\phi) = \pp_\mu +
\Gamma_\mu(\phi)$ are given in~\cite{us},~\cite{us2}, where the logarithmically
divergent contributions have been evaluated.  Explicitly evaluating (2) with an
ultraviolet cut-off $\Lambda$ and a
massive Pauli-Villars sector with a squared mass matrix of the form
$$M_{PV}^2 = H^{PV}(\phi) + \pmatrix{\mu^2& \nu\cr \nu^{\dag}&\mu^2\cr} \equiv
H^{PV} + \mu^2 + \nu, \;\;\;\; |\nu|^2\sim \mu^2\gg H^{PV}\sim H, $$
gives, with $H' = H + H^{PV}$:
\bea 32\pi^2S_1 &=& - \int d^4xd^4p\STr\ln\(p^2 + \mu^2 + H' + \nu\)
+ 32\pi^2\(S'_1 + T_-\)
\nonumber \\
&=& 32\pi^2\(S'_1 + T_-\) - \int d^4xd^4p\STr\ln\(p^2 + \mu^2\) \nonumber \\ &
&
- \int d^4xd^4p\STr\ln\[1 + \(p^2 + \mu^2\)^{-1}\(H' + \nu\)\] . \eea
$S'_1$ is a logarithmically divergent contribution that involves the operator
$\hG_{\mu\nu} = [\hD_\mu,\hD_\nu]$.  Finiteness of (4) requires
\beq  \STr \mu^{2n} = \STr H' = \STr\(2\mu^2H' + \nu^2\) = 0,
\;\;\;\; {1\over64\pi^2}\STr H'^2 = - L,\eeq
where $L$ is the coefficient of ln$\Lambda^2$ in $S'_1 + T_-$.
The vanishing of $\STr \mu^{2n}$ is automatically assured by supersymmetry.
Once the remaining conditions are satisfied we obtain
\beq S_1 = - \int {d^4x\over64\pi^2}\STr\[\(2\mu^2 H' + \nu^2\)\ln\mu^2\]
 + O(\ln\mu^2).\eeq

First consider a supergravity theory in which the Yang-Mills fields have
canonical kinetic energy. Then the quadratically divergent contributions
from the (gauge-fixed) gravity sector, the $N$ chiral supermultiplets and
the (gauge-fixed) Yang-Mills sector of internal symmetry
dimension $N_G$, are respectively:
\bea \STr H_{grav} &=& -10V -2M_\psi^2 - {r\over2} + 4K_{i\m}
\D_\nu z^i\D^\mu\z^{\m} - {x\over 2}F^2, \nonumber \\
\STr H_\chi &=&  2N\(\hV + M_\psi^2 - {r\over4}\) + 2x^{-1}\D_aD_i(T^az)^i
\nonumber \\ & & \qquad
- 2R_{i\m}\(e^{-K}\A^iA^{\m} + \D_\nu z^i\D^\mu\z^{\m}\), \nonumber \\
\STr H_{YM} &=&  2\D + {x\over 2}F^2 + N_G{r\over 2}.\eea
In these expressions, $r$ is the space-time curvature, $F^2 = F^a_{\mu\nu}
F_a^{\mu\nu}$ with $F^a_{\mu\nu}$ the Yang-Mills field strength, $x=g^{-2}$ is
the inverse squared gauge coupling constant, $K_{i\m}$ is
the K\"ahler metric and $R_{i\m}$ is the associated Ricci tensor, $V = \hV
+ \D$ is the classical scalar potential with $\hV = e^{-K}A_i\A^i - 3M_\psi^2
,\;\D = (2x)^{-1}\D^a\D_a, \; \D_a = K_i(T_az)^i$,
and $M_\psi^2 = e^{-K}A\A$ is the field-dependent squared gravitino mass, with
\beq A = e^KW = \A^{\dag}, \;\;\;\; A_i = D_iA, \;\;\;\;
\A^i = K^{i\m}\A_{\m}, \;\;\;\; etc., \eeq
where $D_i$ is the scalar field reparametrization covariant derivative.
The $F^2$ terms in (7) cancel in the overall supertrace.   To regulate the
$z$-dependent terms, where $z=\z^{\dag}$ is the scalar superpartner of a
left-handed fermion, we introduce Pauli-Villars regulator chiral
supermultiplets
$Z_\alpha^I = (\Z_\alpha^{\bar{I}})^{\dag}$, $Z'^I_\alpha =
(\Z'^{\bar{I}}_\alpha)^{\dag}$ and $\varphi^A = (\bv^A)^{\dag}$, with
K\"ahler potential:
\beq K(Z,\Z,\varphi,\bv) = \sum_{\alpha,I=i,M=m}K_{i\m}(z,\z)
\(Z_\alpha^I\Z_\alpha^{\M} + Z'^I_\alpha\Z'^{\M}_\alpha\) +
\sum_A e^{\alpha_A K}\varphi^A\bv^A,\eeq
superpotential:
\beq W(Z,\varphi) = \sum_{A,I}\mu^\alpha_IZ^I_\alpha Z'^I_\alpha
+ \sum_{A}\mu_A\(\varphi^A\)^2, \eeq
and signature $\eta^{\alpha,A} = \pm 1$, which determines the
sign of the corresponding contribution to the supertrace relative to an
ordinary
particle of the same spin.  Thus $\eta = +1 (-1)$ for ordinary particles
(ghosts).  The contribution of Pauli-Villars loops should be regarded as a
parametrization of the result of integrating out heavy ({\it e.g.} Kaluza-Klein
or string) modes of an underlying finite
theory; the contributions from Pauli-Villars fields with negative
signature could be interpreted as those of ghosts corresponding to heavy fields
of higher spin.  $Z^I$ transforms like $z^i$ under the gauge group, and
$Z'^I$ transforms according to the conjugate representation.

In evaluating the effective
one-loop action we set to zero all background
Pauli-Villars fields; then the contribution of these fields to $\STr H_\chi$ is
\bea \STr H^{PV}_\chi &=& 2\sum_{\alpha,A}\(2\eta_\alpha + \eta_A\)
\(\hV + M_\psi^2 - {r\over4}\) \nonumber \\ & &
+ 4\sum_{\alpha,J}\eta_\alpha\[x^{-1}\D^aD_{(\alpha J)}(T_az)^{(\alpha J)} -
R^{(\alpha J)}_{(\alpha J)i\m}\(\A^iA^{\m}e^{-K} + \D_\mu z^i\D^\mu\z^{\m}\)\]
\nonumber \\ & & + 2\sum_A\eta_A\[x^{-1}\D^aD_A(T_az)^A -
R^A_{A i\m}\(\A^iA^{\m}e^{-K} + \D_\mu z^i\D^\mu\z^{\m}\)\]. \eea
 From (9) we obtain for the relevant elements of the
scalar reparametrization connection $\Gamma$ and Riemann tensor $R$:
\bea \Gamma^A_{Bk} &=& \alpha_A\delta^A_BK_k \;\;\;\;
R^A_{B k\m} = \alpha_A\delta^A_B K_{k\m}, \;\;\;\; D_A(T_az)^A = \alpha_A
K_i(T_az)^i, \nonumber \\
\Gamma^{(I\alpha)}_{(J\beta),k} &=& \delta^\alpha_\beta\Gamma^i_{jk},
\;\;\;\; R^{(I\alpha)}_{(J\beta)k\m} =
\delta^\alpha_\beta \delta^I_JR^i_{jk\m}, \;\;\;\; D_I(T_az)^I = D_i(T_az)^i.
\eea
Finally, to regulate the $r$-dependent term in $\STr H$ we introduce
$U(1)$ gauge supermultiplets $W^a$ with signature $\eta^a$ and
chiral multiplets $Z^a = e^{\theta^a} = \(Z^a\)^{\dag} = \(e^{\bth^a}\)^{\dag}$
with the same signature, $U(1)_b$ charge $q_a\delta_{ab}$, and
K\"ahler potential:
\beq
K(\theta,\bth) = {1\over2}\sum_a \nu_a e^{\alpha_a K}(\theta_a + \bth_a)^2\eeq
which is invariant under $U(1)_b$: $\delta_b\theta_a = -\delta_b\bth_a = iq_a
\delta_{ab}$.  The corresponding D-term:
\beq \D(\theta,\bth) = {1\over 2x}\D_\theta^a\D^\theta_a, \;\;\;\;
\D^\theta_a =\sum_bK_b\delta_a
\theta^b = i(\theta^a + \bth^a)q_ae^{\alpha_aK}\nu_a,\eeq
vanishes in the background, but $(\theta^a + \bth^a)/\sqrt{2}$ acquires a
squared mass $\mu_a^2 = (2x)^{-1}q^2_ae^{\alpha_aK}\nu_a$ equal to that of
$W^a$, with which it
forms a massive vector supermultiplet.  The chiral multiplets contribute
to $\STr H^{PV}$ in the same way as $\varphi^A$ with $\alpha_A\to
\alpha_a$, and the vector multiplets contribute to the $r$-term with opposite
sign.  Therefore we obtain an overall contribution from light and heavy modes:
\bea
\STr H' &=&  2\hV\[N\(1+2\sum_\alpha\eta_\alpha\) +
\sum_A\eta_A\(1 - \alpha_A\)
+ \sum_a\eta_a\(1 - \alpha_a\) - 5\] \nonumber \\ & & +
2M_\psi^2\[N\(1+2\sum_\alpha\eta_\alpha\) + \sum_A\eta_A\(1 - 3\alpha_A\)
+ \sum_a\eta_a\(1 - 3\alpha_a\) - 1\] \nonumber \\ & &
- {r\over2}\[N\(1+2\sum_\alpha\eta_\alpha\) + \sum_A\eta_A + 1 - N_G\]
\nonumber \\ & &
+ 2\[x^{-1}\D_aD_i(T^az)^i - R_{i\m}\(e^{-K}\A^iA^{\m} +
\D_\nu z^i\D^\mu\z^{\m}\)\]\(1+2\sum_\alpha\eta_\alpha\)
\nonumber \\ & & + 2\(K_{i\m}\D_\mu z^i\D^\mu\z^{\m} - 2\D\)\(2 -
\sum_A\eta_A\alpha_A - \sum_a\eta_a\alpha_a\).\eea
Thus $\STr H' = 0$ requires
\bea 0 &=& 1+2\sum_\alpha\eta_\alpha = \sum_A\eta_A
+ \sum_a\eta_a - 7 \nonumber \\ &=&
\sum_A\eta_A + 1 - N_G = 2 - \sum_A\eta_A\alpha_A - \sum_a\eta_a\alpha_a.\eea

The vanishing of $\STr(\mu^2 H' + \nu^2)$
further constrains the parameters $\mu(z)$ and $\nu(z,\z)$, which can
in general depend on the light chiral multiplets.  For example, if the
underlying theory is a superstring theory, there is usually invariance under a
modular transformation on the light superfields under which $K\to K+ F(z) +
\bF(\z),\;\;W\to e^{-F(z)}W,$ which cannot be broken by perturbative quantum
corrections~\cite{mod}.  Thus the field $Z^I_\alpha$ has the same modular
weight as $z^i$, $\varphi^A$ has modular weight $-\alpha_A/2$;
the $z$-dependence
of $\mu(z)$, as well as of $\nu_a(z,\z)$ must be chosen so as to restore
modular
invariance.  The result would be interpreted as threshold effects arising from
the integration over heavy modes. (It is possible that some or all of the
modular invariance may be restored by a universal Green-Schwarz type counter
term, as is  the case for the anomalous Yang-Mills
coupling~\cite{dixon}--\cite{tom}.)
In the following we set $q_a = 1,\;\mu_I^\alpha
= \beta_\alpha^Z\mu_I(z), \; \mu_A = \beta_A\mu_\varphi(z), \; \nu_a
= x\beta^2_a|\mu_\theta(z)|^2$, with $\beta_{\alpha,A,a}$ independent of $z$
and $\alpha_{A,a} \equiv \alpha_{\varphi,\theta}$ independent of $A,a$.
Since $\STr(2\mu^2 H' + \nu^2)$ is just the $O(\mu^2)$ part of
$\STr(\mu^2 + H' +
\nu)^2$, it can be read off from the general results of~\cite{us},\cite{us2},
with $H\to H' + \mu^2 + \nu\equiv \tH$. The terms in ${1\over2}\STr \tH$
proportional to $\mu^2$ are:
\bea {1\over2}\STr\tH &\ni& e^{-K}A_{IJ}\A^{IJ}\[\(2\hV + 3M_\psi^2\)
-{r\over2}\] + 2e^{-2K}A_{IJ}\A^{JK}R^{m\;I}_{\;n\;K}A_m\A^n
\nonumber \\ & & + e^{-2K}\[A_{kIJ}\A^{IJm}\A^kA_m
- (A_{IjK}\A^{IK}\A^j A + {\rm h.c.})\] \nonumber \\ & &
 +\D_\mu\z^{\m}\D^\mu z^ie^{-K}\(A_{iJK}\A^{JK}_{\m}
 + 2R^K_{i\m J}e^{-K}A_{KL}\A^{JL}\) \nonumber \\ & &
- {e^{-K}\over x}\D_a(T^az)^iA_{iJK}\A^{JK} \nonumber \\ & &
- {4e^{-K}\over x}\delta^a\theta^c\delta_a\theta^{\bb}\[A_{dc}\A^d_{\bb} -
R^k_{\;n\bb c}A_k\A^n\] +4\(\hV + M_\psi^2\)\K^a_a \nonumber \\ & &
+ {4\over x}\D_\mu z^i\D^\mu\z^{\m}\[R_{\m i\bb c}
\delta_a\theta^c\delta^a\theta^{\bb}
- K_{c\bb}D_{\m}\delta_a\theta^{\bb}
D_j\delta^a\theta^c\].
\eea
where here upper case indices refer to $Z_\alpha^I,Z'^I_\alpha$ and
$\varphi^A$.
Lower indices denote scalar field reparametrization invariant derivatives, and
indices are raised with the inverse metric.
The relevant matrix elements are, in addition to (12):
\bea \sum_Je^{-K}A_{I\alpha,J\alpha}\A^{K\alpha,J\alpha} &=&
\delta^K_I\sum_{m=M}e^K\(K^{i\m}\beta_\alpha^Z\)^2\mu_I\bmu_{\M} \equiv
\delta^K_I\(\beta^Z_\alpha\)^2\Lambda^2_I \nonumber \\
A_{I\alpha,J\alpha,k} &=& \(K_k - \pp_k\ln\mu_I\)A_{I\alpha,J\alpha} -
2\Gamma^{\ell}_{ki}A_{L\alpha,J\alpha}
\nonumber \\
e^{-K}A_{CB}\A^{AC} &=& \delta^A_Be^{K(1-2\alpha_\varphi)}
|\beta_A\mu_\varphi|^2 \equiv \delta^A_B\beta_A^2
\Lambda^2_\varphi \nonumber \\
A_{ABi} &=& \[K_i(1-2\alpha_A) - \pp_i\ln\mu_\varphi\]A_{AB}
\nonumber \eea \bea
\K^a_a &=& {1\over x}\sum_{b,c}\delta^a\theta^cK_{c\bb}\delta_a\bth^{\bb} =
|\beta_a\mu_\theta|^2e^{\alpha_\theta K}\equiv
\beta_a^2\Lambda^2_\theta, \nonumber \\
A_{bc} &=& \nu_be^{\alpha_\theta K} A^{\bb}_c = K_{ab}A - \Gamma^i_{ab}A_i
= \nu_be^{\alpha_\theta K}\[A -\(\alpha_\theta K_{\m} + 2\pp_m\ln\bmu_\theta
\)A^{\m}\],
\nonumber \\
D_i\delta_a\theta^c &=& \Gamma_{ib}^c\delta_a\theta^b = \(\alpha_\theta K_i +
2\pp_i\ln\mu_\theta\)\delta_a\theta^c, \nonumber \eea \beq
\sum_a\delta^a\theta^c\delta_a\theta^{\bb}R_{i\m c\bb}
= -\delta^c_b\beta^2_b\alpha_\theta K_{i\m}\Lambda^2_\theta \eeq
The finiteness constraint requires
\beq\sum_\alpha\eta_\alpha^Z\(\beta^Z_\alpha\)^2 =
\sum_A\eta_A\(\beta_A\)^2 = \sum_a\eta_a\(\beta_a\)^2 = 0.\eeq
Then the results of~\cite{us},~\cite{us2} determine the $O(\mu^2)$
contribution to $S_0 + S_1 = \int d^4x\(\L_0 + \L_1\)$:
\bea \L_0(g_{\mu\nu}^0,K) + \L_1 &=& \L_0(g_{\mu\nu},K + \delta K), \;\;\;\;
g_{\mu\nu} = g_{\mu\nu}^0\(1 + \epsilon\) \nonumber \eea \bea
\epsilon &=& -\lll e^{-K}A_{PQ}\A^{PQ} = \LLL\zeta'_\Phi, \nonumber \\
\delta K &=& \lll\(e^{-K}A_{PQ}\A^{PQ} -4\K_P^P\) = \LLL\zeta_\Phi,\eea
where~\cite{app} $\lambda_\Phi = 2\sum_p\eta^\Phi_p
\(\beta^\Phi_p\)^2\ln\beta^p_\Phi,\;\;\zeta_Z=\zeta_\varphi=\zeta'_Z
=\zeta'_\varphi=1,\;\zeta_\theta = -4\;\zeta_\theta' = 0,$ and $P,Q$ denote all
heavy modes.  It should be emphasized that if there are three or more
terms in the sum over $p$, the sign of $\lambda_\Phi$ is
indeterminate~\cite{app}, so caution should be used in making conclusions
about the implications of these terms for the stability of the effective
potential.

Before proceeding to the case of noncanonical gauge field kinetic energy, we
note that there is an ambiguity in the separation of the fermion loop
contribution into helicity-odd and -even parts.  We define~\cite{us2}:
\bea
-{i\over 2}\Tr\ln(-i\notD + M_\Theta) \equiv -{i\over 2}\Tr\ln\cM(\gamma_5)
 = T_- + T_+, \nonumber \eea
\bea T_- &=& -{i\over 4}\[\Tr\ln\cM(\gamma_5) - \Tr\ln\cM(-\gamma_5)\],
\nonumber \\
T_+ &=& -{i\over 4}\[\Tr\ln\cM(\gamma_5) + \Tr\ln\cM(-\gamma_5)\], \nonumber \\
\cM &=&\gamma_0(-i\notD + M_\Theta) = \pmatrix{\sigma_+^\mu D^+_\mu & M^+\cr
M^-
&\sigma_-^\mu D^-_\mu\cr},\;\;\;\; \sigma_{\pm}^\mu = (1, \pm\vec \sigma).\eea
Thus if $D_\mu = \pp_\mu + V_\mu + iA_\mu\gamma_5,\; M = m + m'\gamma_5$, then
$D_\mu^{\pm} = \pp_\mu + V_\mu \pm iA_\mu,\; M^{\pm} = m \mp m'.$
The ambiguity arises because we can interchange terms that are even and odd in
$\gamma_5$ using $\gamma_5 = (i/24)\epsilon^{\mu\nu\rho\sigma}\gamma_\mu
\gamma_\nu\gamma_\rho\gamma_\sigma$ and similar identities.  In most cases the
correct choice is dictated by gauge or K\"ahler covariance.
However there is an off-diagonal
mass term that mixes gauginos with the auxiliary field $\alpha$:
\beq M_{\alpha\lambda^a} = -\sqrt{x\over2}F_a^{\mu\nu}\sigma_{\mu\nu} =
-\sqrt{x\over2}\(\alpha F_a^{\mu\nu} + i\beta\gamma_5
 \tF_a^{\mu\nu}\)\sigma_{\mu\nu},\;\;\;\; \alpha + \beta = 1.\eeq
The result is invariant under the choice of $\alpha$ only if the integrals are
finite.  In the above we took $\alpha = 1, \;\beta=0$; with an arbitrary choice
we would have gotten, instead of (7):
\bea \STr H_{grav} &=& -10V -2M_\psi^2 - {r\over2} + 4K_{i\m}
\D_\nu z^i\D^\mu\z^{\m} + {x\over 2}F^2(\alpha^2 - \beta^2 -2),
\nonumber \\
\STr H_{YM} &=&  2\D + {x\over 2}F^2(\alpha^2 - \beta^2) + N_G{r\over 2}.\eea
The choice used above is ``supersymmetric'' in the sense that it corresponds to
analogous matrix elements~\cite{us2} in the bosonic and ghost sectors,
yielding the cancellation of the $F^2$ terms.

Now we introduce the dilaton; that is, we couple the Yang-Mills sector to
a holomorphic function of the chiral multiplets:
$f_{ab} = \delta_{ab}f, \;\; f= x+iy$. (The results can
immediately be generalized to the case $f_{ab} = \delta_{ab}k_af,\;k_a=$
constant, by making the substitutions $F^a_{\mu\nu}\to k_a^{1\over2}
F^a_{\mu\nu}, \;A^a_\mu\to k_a^{1\over2}A^a_\mu,\; T^a\to
k_a^{-{1\over2}}T^a.$)
There is a dilatino-gaugino mass term and an additional gaugino connection
that can be written as
\bea M_{\chi^i\lambda^a} &=& -i{f_i\over4\sqrt{x}}\(\gamma F_a^{\mu\nu}
+ i\delta\gamma_5\tF_a^{\mu\nu}\)\sigma_{\mu\nu},
\;\;\;\; \gamma + \delta = 1, \;\;\;\; f_i = \pp_if \nonumber \\
A^\mu_{\lambda^a\lambda^b} &=& -\delta_{ab}{\pp^\mu
y\over2x}\(i\epsilon\gamma_5
- \zeta{\epsilon^{\lambda\nu\rho\sigma}\over24}\gamma_\lambda
\gamma_\nu\gamma_\rho\gamma_\sigma\), \;\;\;\; \epsilon + \zeta = 1.\eea
Then we obtain the additional contributions to the supertraces:
\bea
\STr H_{YM} &\ni&  {f_i\f^i\over 4x}F^2(\gamma^2 - \delta^2) -
N_G\(2M_\lambda^2
+ {1\over2x^2}\[\pp_\mu x\pp^\mu x + (3 - 2\zeta^2)\pp_\mu y\pp^\mu y\]\),
\nonumber \\ \STr H_{grav} &\ni& {f_i\f^i\over 4x}F^2(\gamma^2 - \delta^2)
 - {f_i\f^i\over2x^2}\D,
\;\;\;\; \STr H_\chi \ni {f_i\f^i\over2x^2}\D, \eea
where $M_\lambda^2 = (2x)^{-2}e^{-K}f_i\f^jA_j\A^i,\; \f^i = K^{i\m}\f_{\m}.$
The
``supersymmetric'' choice, which matches corresponding matrix
elements~\cite{us2} in the
bosonic and ghost sectors, is $\gamma=\delta={1\over2},\;\epsilon=0,\;\zeta=1.$
Then the $F^2$ terms again cancel, and the remaining terms:
\beq \STr H \ni - N_G\(2M_\lambda^2
+ {1\over2x^2}\[\pp_\mu x\pp^\mu x + \pp_\mu y\pp^\mu y\]\),\eeq
can be regulated by the introduction of additional Pauli-Villars chiral
multiplets, as will be shown below.  With any other choice cancellation of the
infinities would be achieved only through the
introduction of Pauli-Villars ``dilatons'' and/or ``gauge fields'' with linear
couplings to the light, physical fields, thus entailing
loops mixing quantum fields of different signature.  With the choice
$\zeta = 1$, the $y$-axion contribution
to the gaugino connection can be written as
\beq A_\mu = - {x\over3}h^{\nu\rho\sigma}
\gamma_{[\mu}\gamma_\nu\gamma_\rho\gamma_{\sigma]},\eeq where
$h^{\nu\rho\sigma}= 4x^2\epsilon^{\nu\rho\sigma\mu}\pp_\mu y$ is the
3-form
that is dual to the axion in absence of interactions.  The axion also appears
through the 3-form in a contribution to the gauge boson connection~\cite{us2}.
This
suggests that the linear supermultiplet formalism~\cite{linear} is the natural
framework for describing the dilaton supermultiplet, at least in the absence of
a superpotential for the dilaton. It has been shown~\cite{frad} that the
axion/3-form duality holds at the quantum level,
up to finite topological anomalies.  Here we see that when couplings to
fermions
are included, there are additional anomalies; shifting contributions between
$T_+$ and $T_-$ in (21) is analogous to shifting the integration variable in a
Feynman diagram calculation.  For example,
the linearly divergent triangle diagram leads to
an ill-defined finite chiral anomaly that is fixed by imposing invariance under
local gauge transformations.  In the present case supersymmetry must be used to
resolve the ambiguity.  With the choice $\zeta=1$, the gaugino connection (24)
is purely ``vector-like'', and does not contribute to the anomalous $F\tF$
term that breaks~\cite{anomalies} modular invariance, and, by
construction~\cite{us2}, is contained in $T_-$.  This
agrees with the conclusions of~\cite{tom}, where it was argued that such a
contribution is inconsistent with the linearity constraint in the linear
multiplet formulation.

To regulate the terms in (26) we introduce chiral supermultiplets
$\pi^\alpha = \(\bar{\pi}^\alpha\)^{\dag}$  with
\beq K(\pi,\bar{\pi}) = \sum_\alpha(f + \f)|\pi^\alpha|^2, \;\;\;\;
W(\pi) = \sum_\alpha\beta_\pi^\alpha \mu_\pi(z)(\pi^\alpha)^2, \;\;\;\;
\eta^\pi_\alpha = \pm 1.\eeq  Then
\beq \Gamma^\beta_{\alpha i} = {f_i\over 2x}\delta^\beta_\alpha, \;\;\;\;
R^\beta_{\alpha i\m} = - {f_i\f_{\m}\over4x^2}\delta^\beta_\alpha, \;\;\;\;
D_\alpha(T_az)^\alpha = {f_i\over2x}(T_az)^i = 0.\eeq
Inserting this into the general expression (7) for $\STr H_\chi$ we get a
contribution $\STr H^\pi$ that cancels (26) provided
$\sum_\alpha\eta^\pi_\alpha
= + N_G$, and the conditions (16) are modified accordingly.  Letting $I
,J,\cdots $ denote also $\pi_\alpha$ in (17), we get additional
contributions to STr$(2\mu^2 H' + \nu^2)$:
\bea
e^{-K}A_{\gamma\beta}\A^{\alpha\gamma} &=& {e^K\over4x^2}
\delta_\beta^\alpha |\beta_\alpha^\pi\mu_\pi|^2 \equiv
\delta_\beta^\alpha\(\beta_\alpha^\pi\)^2\Lambda^2_\pi,
\nonumber \\
A_{\alpha\beta i} &=& \[K_i - {f_i\over x} - \pp_i\ln\mu_\pi\]A_{\alpha\beta}.
\eea
Including these, with $\sum_\alpha\eta_\alpha^\pi\(\beta_\pi^\alpha\)^2 = 0$,
(20) is modified to include $P= \pi_\alpha,\;\Phi= \pi,\;\zeta_\pi =
\zeta'_\pi = 1$.

To fully regulate the theory, including all logarithmic divergences, additional
Pauli-Villars fields and/or couplings must be included.  Specifically, the
superpotential must include the terms
\beq W\ni {1\over 3}\sum_\alpha W_{ij}Z^I_\alpha Z^J_\alpha, \eeq
and, to regulate the Yang-Mills contributions, we must include in the set
$\varphi^A$
chiral multiplets $\varphi^A_a = (\bv^A_a)^{\dag}$ that transform according to
the adjoint representation of the gauge group, with
$\sum_{A,a}\eta^a_A = 3N_G$.
The field dependence of the corresponding effective cut-off was
determined in~\cite{tom} by imposing the supersymmetric relation between the
chiral and conformal anomalies.  This in turn determines the K\"ahler
potential: $\alpha_A^a = {1\over 3}$.  Imposing the full finiteness condition
on
$\STr H'^2$ may constrain the other parameters $\alpha_A,\alpha_a,
\alpha_\alpha.$

\vskip .3in
\noindent{\bf Acknowledgements.} I thank Jon Bagger and Soo-Jong Rey for
discussions that prompted completion of this work, most of which was done at
the Aspen Center for Physics.  This work was supported in part by the
Director, Office of Energy Research, Office of High Energy and Nuclear Physics,
Division of High Energy Physics of the U.S. Department of Energy under Contract
DE-AC03-76SF00098 and in part by the National Science Foundation under grant
PHY-90-21139.

\end{document}